\documentstyle[11pt,newpasp,twoside]{article}
\def\edcomment#1{\iffalse\marginpar{\raggedright\sl#1\/}\else\relax\fi}
\reversemarginpar

\begin{document}
\title{Theoretical predictions for the cold part of the colliding  
       wind interaction zone}
\author{Doris Folini}
\affil{Observatoire de Strasbourg, 67000 Strasbourg, France}
\author{Rolf Walder}
\affil{Institute of Astronomy, ETH Z\"{u}rich, 8092 Z\"{u}rich, Switzerland}

\begin{abstract}

We present 2D and 3D hydrodynamical simulations of the colliding wind
interaction zone in WR+O binaries. It is shown that 3D effects can
basically explain certain observed, orbit dependent flux
variations. Possible connections between the interior structure of the
interaction zone and dust formation are outlined and its stability is
re-investigated.

\end{abstract}

\section{Introduction}
\label{sec:intro}

A sufficient theoretical understanding of the wind-wind collision
zone in WR+O binaries is essential to achieve a physically correct
interpretation of the observational data.  The task is difficult as
many physical processes and scales are involved and as, at least for
some phenomena, 3D studies are unavoidable.

The presented results are based on numerical simulations. Bearing in
mind the character of the workshop, we try to bring them together with
thoughts and contributions of other participants.  We address the
global shape of the interaction zone in Section~2. Its interior is the
topic in Section~3, and Section~4 deals with stability. Conclusions
follow in Section~5.

\begin{figure}[tb]
\centerline{
   \plotfiddle{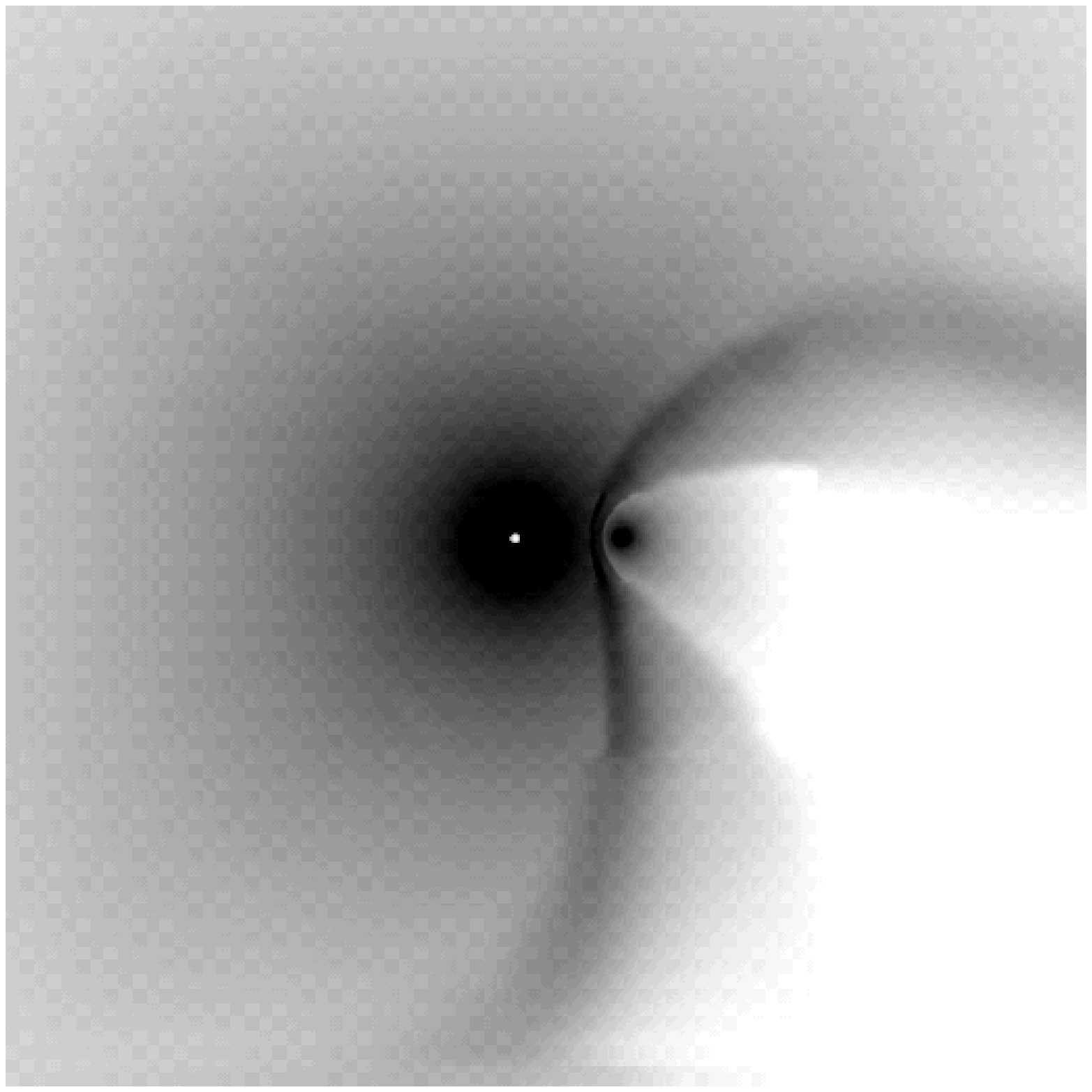}{5.8cm}{0}{28}{28}{-175}{-40}
   \plotfiddle{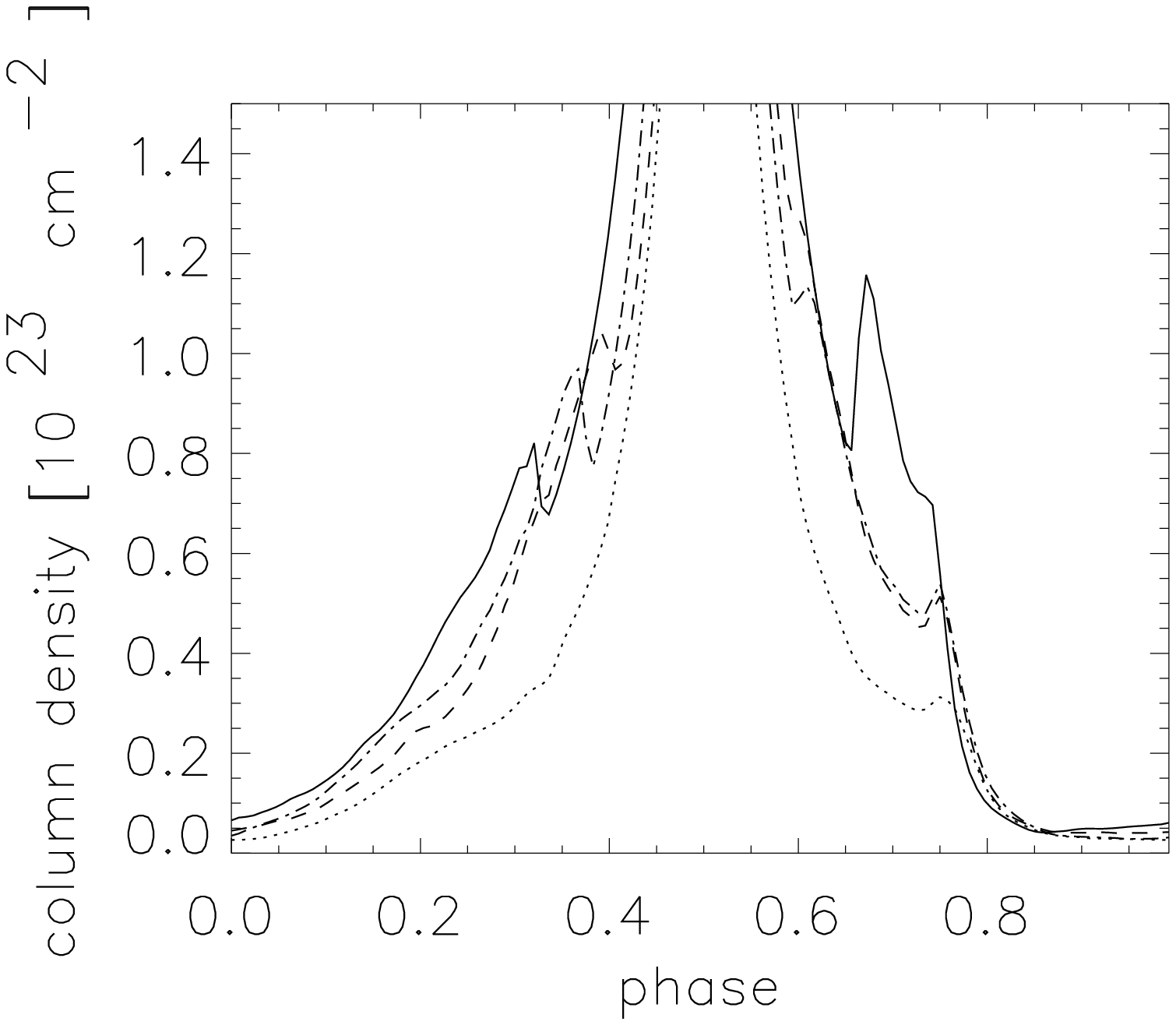}{5.8cm}{0}{40}{45}{-420}{-40}
           }
\caption{Spirally shaped colliding wind interaction zone and resulting
         column density. {\bf Left:} Density distribution in the
         orbital plane at periastron in an adiabatic 3D simulation of
         $\gamma$ Velorum (logarithmic scale, high density is black,
         WR-star (white) to the left, O-star (black) to the
         right). The x- and y-extension of the slice shown are
         $10^{14}$cm. {\bf Right:} Variation of the column density,
         measured along $ 1.9 \cdot 10^{14}$cm long rays towards the
         O-star. Each individual curve denotes the column density an
         observer would see if, at one fixed time, he were to move
         around the system within the orbital plane. The four
         different curves denote four different times.  {\it Solid
         line:} periastron. {\it Dashed line:} quadrature following
         periastron. {\it Dotted line:} apastron. {\it Dash-dotted
         line:} quadrature following apastron. X-axis: location of the
         observer (0: observer on the line connecting WR- and O-star,
         on the side of the O-star). Increasing x: observer proceeds
         in direction of orbital motion (counter clockwise in the left
         picture). Y-axis: column density in units
         $10^{23}$cm$^{-2}$. Note the asymmetry with respect to
         x=0.5. The chopped, central peak corresponds to the
         undisturbed WR-wind. }
\label{fig:spiral}
\end{figure}
\section{A spiral in 3D}
\label{sec:globstru}

As the two stars orbit around each other, the O-star carves a spirally
shaped 'tunnel' out of the WR-material. The O-star wind material is
separated from the WR-wind material by the wind-wind interaction zone.
In the more central part of the system, this zone is confined by
shocks. There, wind material gets heated and compressed as it enters
the interaction zone.  Far enough away from the center, such shocks
confining the interaction zone are essentially absent. There, no
further heating or compression occurs. Instead, all matter moves
outwards at approximately the same speed and the high density
interaction zone disperses.

In the observers frame, the spiral just described can be regarded as a
wave pattern as it is not linked to any spirally shaped particle
paths. In this frame, particles essentially proceed radially outwards
with respect to the system center. In circular systems the spiral
pattern is stationary in the frame co-rotating with the system. In
highly eccentric systems, on the other hand, the spiral pattern
(e.g. opening angle of the spiral) undergoes slight changes as a
function of orbit (for details see Walder, Folini \& Motamen 1999, WFM
in the following).

This spirally shaped matter distribution has observable consequences.
For the case of $\gamma$ Velorum it was shown in WFM that with such a
spirally shaped interaction zone the observed asymmetric X-ray light
curve (Willis, Schild, \& Stevens 1995) can basically be
understood. We predict that such a matter distribution can also
account for orbit dependent line flux variations as observed in GP Cep
(see Demers, this volume) and other systems.

As we have not modeled any of these systems in great detail so far, we
build our argumentation on a related, however adiabatic model instead.
The presented results, therefore, have qualitative value
only. Figure~\ref{fig:spiral} shows the spirally shaped density
distribution in the orbital plane of an adiabatic 3D simulation of the
system $\gamma$ Velorum, along with derived column densities towards
the O-star. For details of the 3D simulation see WFM. As can be taken
from this figure, considerably enhanced column densities, associated
with the spirally shaped interaction zone, do occur in this case. The
enhancement is more pronounced for the trailing edge of the spiral
than for its leading edge. And the enhancement does not occur all the
time. While being rather pronounced at periastron, hardly any
enhancement is present at apastron. This difference is due to the
highly eccentric orbit of $\gamma$ Velorum. For another system than
$\gamma$ Velorum, with other wind parameters, the quantitative values
will be different.

We predict that these adiabatic results essentially carry over to the
case where the interaction zone efficiently cools close to the center,
i.e. to narrow binaries. Compared to the adiabatic case, the density
increases while the zone becomes geometrically thiner. The resulting
high density interaction zone is located close to where the contact
discontinuity resides in the adiabatic case. If efficient cooling
occurs already in the system center the interaction zone most likely
is unstable. This again increases its apparent geometrical thickness
somewhat while it decreases the column density. In summary, we make on
this basis the following predictions for the column density in the
radiative case. The enhancement is going to be larger but it is going
to cover a shorter phase. The spiral carved by the O-star is going to
be more narrow. Compared to the adiabatic case, the highest column
densities are going to occur at phases further away from x=0.5 in the
notation of Figure~\ref{fig:spiral}, when the trailing edge of the
spiral is in the line of sight towards the observer.

One may speculate that such high column densities in principle can
shield a portion of matter from the intense stellar radiation, thus
promoting dust formation (episodically, as e.g. in WR 137, Marchenko,
Moffat, \& Grosdidier 1999, or permanently, as e.g. in WR 104,
Tuthill, Monnier, \& Danchi 1999). However, more detailed studies are
necessary to decide whether such shielding is efficient enough and can
last long enough to really promote dust formation.

\section{Interior structure}
\label{sec:intstru}

The structure within the cold, high density interior of a radiatively
cooling interaction zone, its density, velocity, temperature, and
chemical composition, are probably of importance with regard to dust
formation and observed large line widths (e.g. L\'{e}pine, Eversberg,
\& Moffat 1999).

The density and velocity in a {\it radiatively cooling interaction
zone} most probably are inhomogeneous for two reasons. First, it seems
clear today that at least the WR-wind is clumped already (Owocki,
Castor, \& Rybicki 1988; Cherepashchuk 1990; L\'{e}pine, Eversberg, \&
Moffat 1999). The interaction zone then is probably clumped at least
as much (see also Walder \& Folini, this volume). Second, even for the
collision of homogeneous flows, 2D plane parallel simulations
including radiative cooling show a highly non-homogeneous, turbulent
density- and velocity-distribution within the interaction zone
(e.g. Blondin \& Marks 1996; Folini \& Walder 2001).

The prediction of the temperature within the high density interaction
zone is a more difficult case (see Folini \& Walder 2000), but low
temperatures must be reachable as dust is observed in some WR+O
binaries.

The chemical composition is particularly important with regard to dust
formation (Le Teuff, this volume; Cherchneff et al. 2000). A central
question here is whether hydrogen rich O-star material and, in dust
forming binaries, carbon rich WR-material can mix efficiently
enough. Mixing itself is a diffusive process and thus not described by
the models presented here. These models do not contain any explicit
physical diffusion. On the other hand, the presented models do contain
numerical diffusion, but this diffusion is not controllable. Other
studies are required to investigate whether physical mixing does occur
fast and efficiently enough to be of importance with regard to dust
formation. What we can say is that the conditions for mixing are
improved in such interaction zones by efficient stirring, due to
turbulence and instabilities (see below).

\section{Stability}
\label{sec:stab}

\begin{figure}[tb]
\centerline{
   \plotfiddle{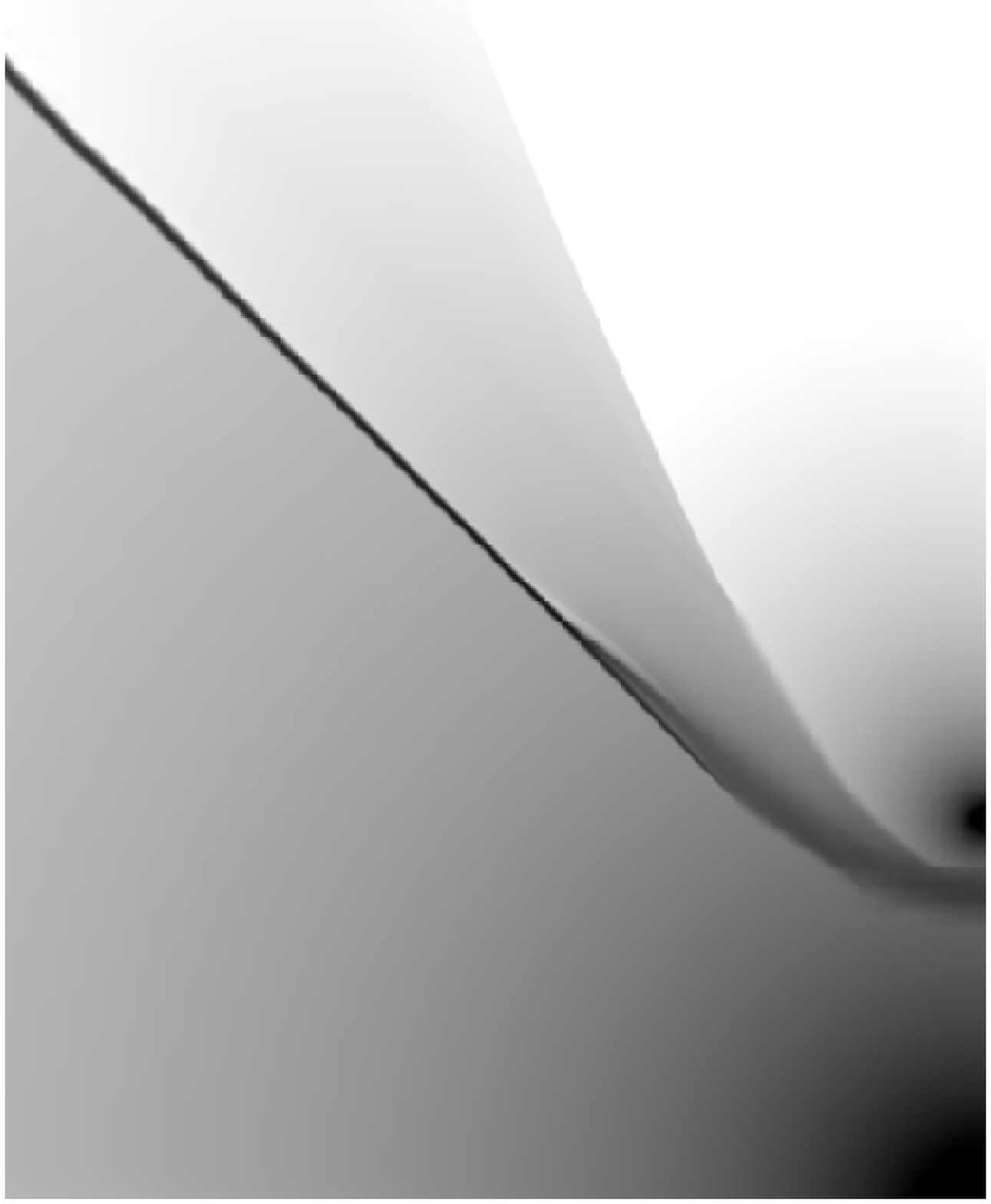}{7.8cm}{0}{36}{32}{-200}{-35}
   \plotfiddle{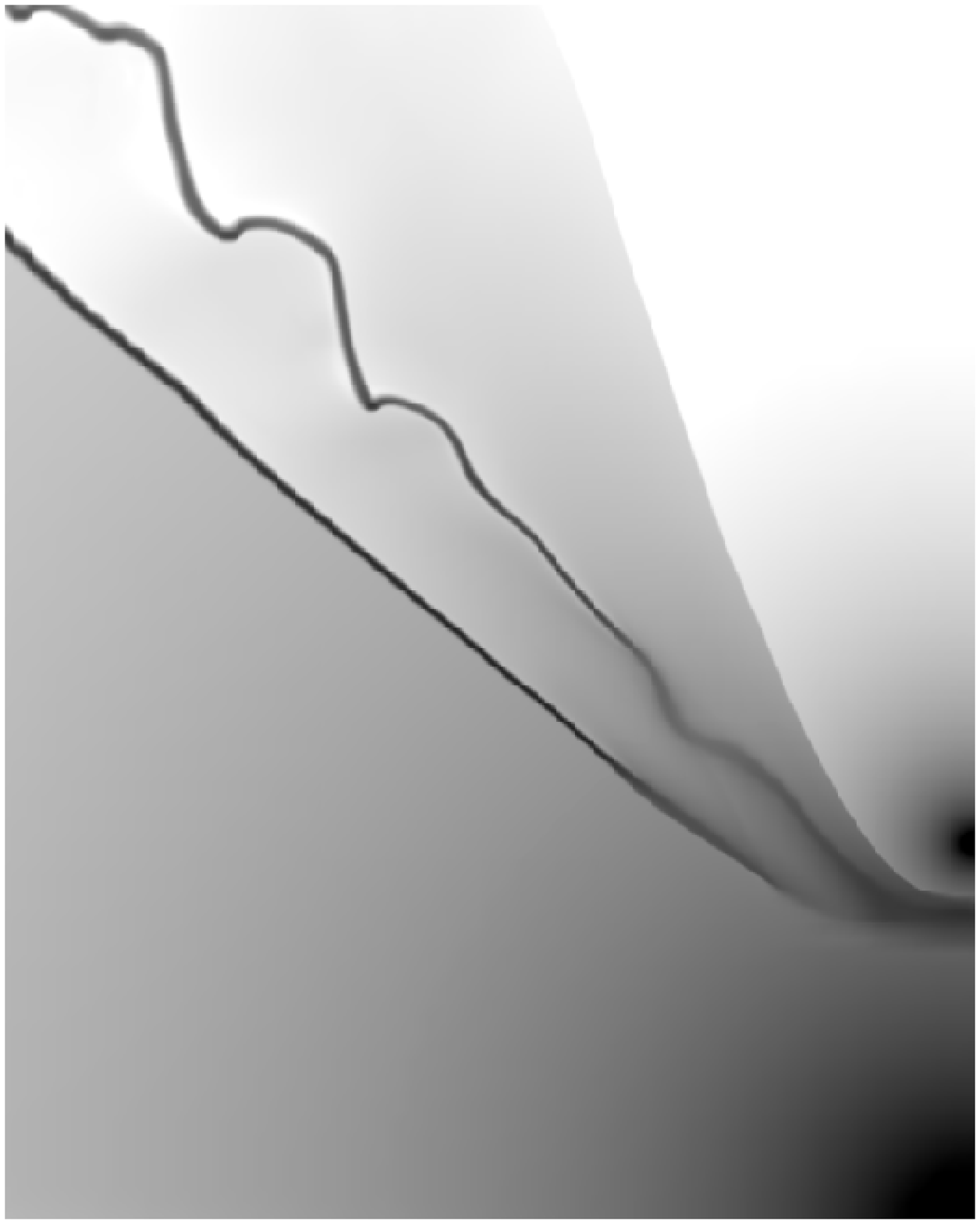}{7.8cm}{0}{33}{32}{-380}{-35}
           }
\caption{2D radiatively cooling simulation of WR 140, shown in density
         (logarithmic scale, high density is black, WR-star at bottom,
         O-star further up, the x-dimension shown is about $7 \cdot
         10^{13}$cm). For the parameters see Walder \& Folini (this
         volume). {\bf Left:} No disturbances except for numerical
         noise are present. The shocked WR wind cools roughly one
         separation away from the system center (dark line). The
         situation is stationary. {\bf Right:} Starting from the left
         figure, a high density knot (3.333... times higher density
         than ambient wind medium, about 2 R$_{\odot}$ big) was
         injected into the interaction zone, about 1/6 separation away
         from the stagnation line.  A cold, high density ribbon forms
         which flutters in the hot, shocked O-star wind. }
\label{fig:stability}
\end{figure}
As a great number of publications already exists on this topic (see
e.g.  Walder \& Folini 1998 for a recent review) we address here only
a few selected issues. 

Let us look at the colliding wind interaction zone in a relatively
wide WR+O binary, assuming that the stellar winds are homogeneous. In
this case, radiative cooling becomes efficient only some distance away
from the system center. Figure~\ref{fig:stability} (left) shows that
in this case the interaction zone can be stationary, provided that
there are no disturbances or at least that disturbances are dispersed
by sound waves before they reach regions of strong radiative
cooling. In particular, the numerical noise, which is always present,
is insufficient to cause instability under these conditions.

But also a single, high density clump injected into such an
interaction zone is probably not enough to cause the interaction zone
as a whole to become unstable.  Figure~\ref{fig:stability} (right)
shows a first attempt to investigate such a scenario. Although we do
not yet fully understand the numerical results we obtain, and not
controllable numerical effects, in particular numerical diffusion,
certainly influence the results, we decided to publish them here to
allow others to think and discuss about them. What we basically
observe is that the injection of such a clump first causes some
shaking of the interface confining the interaction zone, which later
on dies away again. Meanwhile, an additional, cold, high density
ribbon forms, which flutters in the ambient, hot, shocked O-star
material. The starting point of this ribbon stays more or less fixed
in space, close to the system center. The ribbon persists for at least
$10^{8}$ s, much longer than a typical advection time scale in this
system.

Several question arise in connection with this result. With regard to
the limited space, we address only two of them here. First, why does
the ribbon with its starting point close to the system center persist?
An explanation may be numerical diffusion and associated conduction
effects (we have no physical heat conduction included in the presented
models), in combination with radiative cooling. Conduction and cooling
time scales could be such that their effect exactly compensates the
outward transport, leaving the starting point of the ribbon
stationary.  Also compression or ablation effects, as the 'normal'
post shock flow hits the existing ribbon, and subsequent enhanced
cooling could contribute to the persistency of the ribbon. Second, is
the fluttering of the ribbon real? This question is of importance as
it is probably related to the more general question of the stability
of a radiatively cooling wind-wind interaction zone.  Kelvin-Helmholtz
instabilities could be an explanation for the fluttering. If so, the
fluttering would depend on the cooling cut-off used in the radiative
cooling model, which determines the density ratio between the high
density sheet and the surrounding medium. The higher this density
contrast, the smaller the wavelength of the Kelvin-Helmholtz
instability and its growth rate. This seems in agreement with the
findings of Myasnikov, Zhekov, \& Belov (1998), who report a similar
dependence of the stability of the cold, high density interaction zone
in colliding wind binaries on the applied cooling cut-off.

To obtain an unstable, high density interaction zone numerical
simulations suggest that it is necessary to have either a narrow WR+O
binary or to inject a large number of clumps into an otherwise
stationary interaction zone (see Walder \& Folini, this volume).  As
these results can be qualitatively corroborated by physical arguments
we consider them to be {\it qualitatively} reliable. This despite the
fact that there are a number of difficulties associated with {\it
quantitative} numerical results. We only want to mention here one of
the more prominent problems that occur, namely the pile up of mass
along the stagnation line. This carbuncle phenomenon (Quirk 1994) can
be remedied by introducing sufficient artificial viscosity, but other
features may be 'drowned' in this way as well.

\section{Conclusions}
\label{sec:conc}

3D numerical simulations show that the global, spirally shaped matter
distribution in colliding wind binaries can basically account for
observed, orbit dependent flux variations. We suggest that such
shielding effects can also promote the formation of dust. The interior
structure of a cold, high density interaction zone most likely is
inhomogeneous in density and velocity. Efficient stirring of the
WR-star and O-star material is most likely. How much mixing ---
important with regard to dust formation and in contrast to stirring a
diffusive process --- is induced in this way remains to be
investigated. However, 2D simulations show that the cold, high density
interaction zone can be stationary, provided that efficient cooling is
present only some distance away from the system center, i.e. in wide
binaries, and that there are no significant disturbances, like
e.g. clumped winds. These preliminary results also suggest that the
stability of the interaction zone as a whole is not affected by the
injection of one single, high density clump, but that only a cold,
high density ribbon forms within the hot, shocked O-star material.

\acknowledgments These proceedings benefited greatly from discussions
with Sergey Marchenko and Hugues Demers, and from extended biking with
Andy Pollock.


\begin{references}
\reference Blondin, J. M., \& Marks, B. S. 1996, 
           New Astronomy, 1, 3, 235
\reference Cherchneff, I., Le Teuff, Y. H., Williams, P. M., \& 
           Tielens, A. G. G. M. 2000, \aap, 357, 572
\reference Cherepashchuk, A. M. 1990, \sovast, 34, 481
\reference Folini, D., \& Walder, R. 2000,
           in ASP Conference Series 204, 
           Thermal and Ionization Aspects of Flows from Hot Stars:
           Observations and Theory,
           ed. H.~J.~G.~L.~M. Lamers and A. Sapar, 267
\reference Folini, D., \& Walder, R. 2001, submitted to \aap
\reference L\'{e}pine, S., Eversberg, T., \& Moffat, A. F. J. 1999,
           \aj, 117, 1441
\reference Marchenko, S. V, Moffat, A. F. J., \& Grosdidier, Y. 1999,
           \apj, 522, 433
\reference Myasnikov, A. V., Zhekov, S. A., \& Belov, N. A. 1998, 
           \mnras, 298, 1021
\reference Owocki, S.~P., Castor, J. I., \& Rybicki, G.~B. 1988,
           \apj, 335, 914
\reference Quirk, J. J. 1994,
           International J. Numer. Methods Fluids, 18, 555
\reference Tuthill, P. G., Monnier, J. D., \& Danchi, W. C. 1999, 
           Nature, 398, 487
\reference Walder, R. \& Folini, D. 1998, \apss, 260, 215
\reference Walder, R., Folini, D., \& Motamen, S. M. 1999, 
                      in  Proc. IAU Symposium No. 193,
                      Wolf-Rayet Phenomena in Massive Stars and 
                      Starburst Galaxies, 
                      ed. K. A. van der Hucht, G. Koenigsberger, \& 
                               P. R. J. Eenens
                      (ASP), 298 
\reference Willis, A. J., Schild, H., \& Stevens, I. R. 1995, \aap,298, 549
\end{references}
\end{document}